%% file: JBenton_charm2013.tex
\documentclass[12pt]{article}
\usepackage{graphicx}
\usepackage{sidecap}
\usepackage{amsmath}

\def\pbnr{}
\def\speaker{Jack Benton}
\def\onbehalfof{the university of bristol}
\def\title{Finding an Amplitude Model for $D^{0} \rightarrow \pi^{-}\pi^{+}\pi^{+}\pi^{-}$ using CLEO-c Data}
\def\affiliation{School of Physics\\
The University of Bristol, Bristol, UK}
\def\support{The workshop was supported by the University of Manchester, IPPP, STFC, and IOP}

\input charmmacros.tex

\begin{document}
\begin{titlepage}
\pubblock

\vfill
\Title{\title}
\vfill
\Author{\speaker\SupportedBy{\support}\OnBehalf{\onbehalfof}}
\Address{\affiliation}
\vfill
\begin{Abstract}
Amplitude analyses of four body D decays are of significant interest for multiple reasons, including their potential contribution to CP violation studies in both B and D decays. The resonant substructure of a four body decay has many possible components and combinations. Here, a genetic algorithm for the optimisation of such amplitude models is described. The application to an amplitude analysis of the decay mode $D^{0} \rightarrow \pi^{-}\pi^{+}\pi^{+}\pi^{-}$ using CLEO-c data is discussed and the performance of the algorithm is verified on simulated data, showing convergence to the original model.

\end{Abstract}
\vfill
\begin{Presented}
\venue
\end{Presented}
\vfill
\end{titlepage}
\def\thefootnote{\fnsymbol{footnote}}
\setcounter{footnote}{0}
%

\section{Introduction}
Studies of $D$ decays such as $D^{0} \rightarrow \pi^{-}\pi^{+}\pi^{+}\pi^{-}$ contribute to our understanding of low energy QCD interactions and are of high interest for possible CP violation studies. Direct CP violation can be observed by comparing the intermediate resonance structure of $D^{0}$ and $\overline{D^{0}}$, while amplitude models of these decays can be used to improve sensitivity to the unitarity triangle phase $\gamma$ in analyses of $B \rightarrow D K$ and related decays.

Searches for direct CP violation in $D \rightarrow \pi^{-}\pi^{+}\pi^{+}\pi^{-}$ decays have recently been reported~\cite{MattMiranda}, however a CP violation search using an amplitude analysis can help with the interpretation of any CPV signal. They can also provide a more detailed description of CP violation from individual resonances, which may have been masked in other searches.

Measurements of $\gamma$ using amplitude analyses of three-body $D$ decays from $B \rightarrow DK$ have been successful in several examples~\cite{BtoDKgamma1,BtoDKgamma2,BtoDKgamma3}. However, studies suggest that four-body $D$ decays with an amplitude model, such as those described here, could produce competing measurements and significantly contribute to the worldwide average measurement of $\gamma$~\cite{gafrom4body}. An amplitude analysis of $D^{0} \rightarrow \pi^{-}\pi^{+}\pi^{+}\pi^{-}$ has been reported before by the FOCUS collaboration~\cite{FOCUS}, but more data is now available from the CLEO-c detector.

Amplitude analyses for four body decays are challenging due to the many possible intermediate resonant components and the five dimensional phase space of the decay. Here we report on a method to systematically build and refine amplitude models and discuss its application to $D^{0} \rightarrow \pi^{-}\pi^{+}\pi^{+}\pi^{-}$ events from CLEO-c data.

\section{Selection of CLEO-c Data}
The data used in this analysis was produced in symmetric $e^{+}e^{-}$ collisions at the Cornell Electron Storage Ring (CESR) and detected by the CLEO-c experiment from 2003 to 2008, with a total integrated luminosity of 600~fb$^{-1}$. During this period, CESR ran at energies close to the $\psi(3770)$ resonance that decays predominately to $D^{0} - \overline{D^{0}}$ pairs. The CLEO-c detector is built cylindrically outward, starting closest to the beam line with a low mass inner drift chamber for tracking, then a second outer drift chamber, a Ring Imaging CHerenkov detector (RICH) and finally a 7800 crystal CsI electromagnetic calorimeter. Particle identification is provided by measured energy loss and the combination of momentum information with velocity measurements from the RICH.

$D^{0}$ and $\overline{D^{0}}$ separation is achieved by identifying individual charged kaons from the `other side' $D$ decay. Assuming these kaons are a result of Cabbibo-favoured $D$ decays, allows us to `tag' the flavour of both $D$'s with a 95.5\% accuracy~\cite{KKpipiPaper}. It was also found that selecting higher momentum kaons decreases the chance of kaon-pion misidentification and therefore improves tagging accuracy. Candidate $D\rightarrow\pi^{-}\pi^{+}\pi^{+}\pi^{-}$ events must therefore have an `other side' kaon with momentum $>$ 400MeV$/c$.

Signal selection is performed by using the standard CLEO-c selection criteria as described in Ref.~\cite{S_Dobbsetal} on the candidate tracks.

\subsection{Signal and Background Regions}
\label{subsec:SignalandBackgroundRegions}
\begin{figure}[htb]
\centering
\includegraphics[trim = 0mm 0mm 0mm 8mm, clip, scale = 0.5]{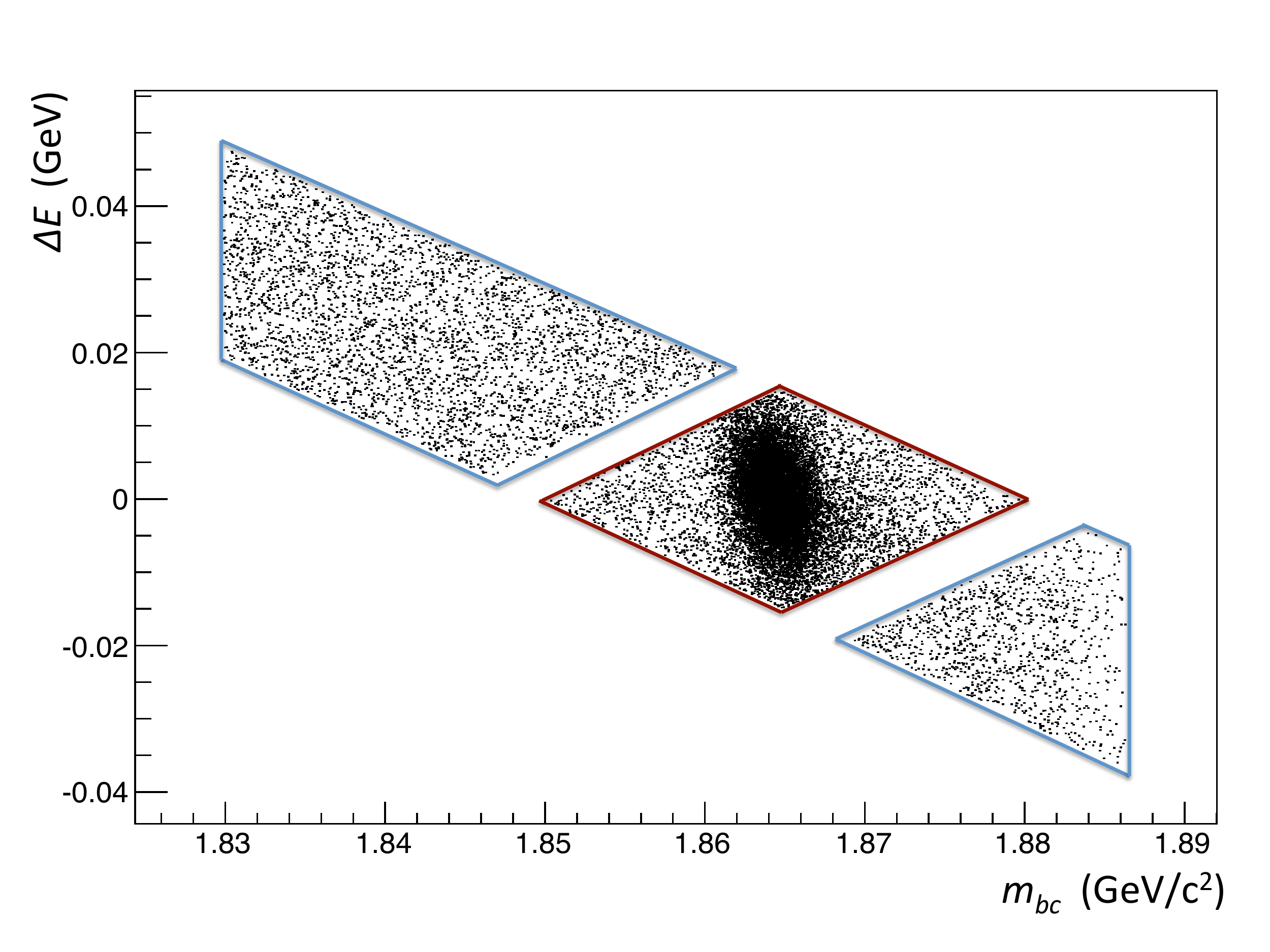}
\caption{The distribution of events in missing energy $\Delta E$ and beam constrained mass within the selection regions. The central region (red) is defined as the signal region, with sideband regions (blue) providing background samples.
}
\label{fig:CutRegions}
\end{figure}

Two kinematic variables are used to define a signal and two `side band' background regions. The variables are defined as: the beam-constrained mass,
\begin{equation}
m_{bc} \equiv \sqrt{\frac{E_{b}^{2}}{4c^{4}}-\frac{\mathbf{p}_{D}^{2}}{c^{2}}},
\end{equation}
where $E_{b}$ is the total energy delivered by the beam in the centre-of-mass frame and $\mathbf{p}_{D}$ is the reconstructed four momentum of the candidate $D$; and $\Delta E$,
\begin{equation}
\Delta E \equiv E_{D} - \frac{E_{b}}{2},
\end{equation}
where $E_{D}$ is the total reconstructed energy of candidate $D$. Signal events should have missing energy ($\Delta E$) close to zero and beam constrained mass close to that of the $D^{0}$ mass.
Amplitude analyses fit probability density functions (P.D.F.s) to the data, within specific invariant mass windows. Different regions of available invariant mass can host different physical processes and distributions. Therefore, it is important to select signal and background regions with a mutual and constant invariant mass, i.e. that of the $D$.
Using these variables we can draw lines of constant invariant mass, such as the $D^{0}$ mass, with the relation:
\begin{equation}
\pm\sqrt{\Delta E^{2} + 2\Delta E E_{b} + m_{bc}^{2}} = m_{D},
\end{equation}
which describes a circle in $m_{bc}$ and $\Delta E$ space. Lines normal to this curve can be described simply by an angle $\theta$ around the centre of this circle, with 
\begin{equation}
\theta \equiv \arctan\left(\frac{\Delta E + E_{b}}{m_{bc}}\right).
\end{equation}
If we then draw a box of sides $\Delta S = m_{D}~\pm$~15 MeV and with $\Delta \theta = \theta_{D} \pm 0.004$, where $\Delta E = 0$ and $m_{bc} = m_{D}$ at $\theta_{D}$, we can define a signal region around the $D$ mass peak, as shown in Figure \ref{fig:CutRegions}. Side band regions are then simply defined with different choices of $\Delta\theta$.

\section{Amplitude Analysis and Techniques}
Amplitude analyses aim to reconstruct the intermediate processes between some initial and observed final states, which must be achieved with only the kinematic information of the final state particles.  One approach to an amplitude analysis, is to treat each intermediate resonance as a Breit-Wigner function~\cite{BreitWigner} of known mass and width. These are then multiplied by a unknown magnitude and complex phase. When these functions, in addition to a non-resonant component, are summed together they form the decay amplitude $\mathcal{M}$ for resonance $R$; 

\begin{equation}
\label{equation:Sum}
\mathcal{M} = a_{0}e^{i\delta_{0}}+\displaystyle\sum_{R} a_{R}e^{i\delta_R}F_{L}^{R}B_{R}F_{L}^{D}M_{L},
\end{equation} where $a_{R}$ and $\delta_{R}$ are the magnitude and phase for each resonant component, $B_{R}$ is the Breit-Wigner function, $F_{L}^{R}$ and $F_{L}^{D}$ are the Blatt-Weisskopf barrier factors for the resonance and $D^0$ respectively, while $M_{L}$ describes the angular distribution of the final state particles~\cite{PhysRev}. 

The constructed amplitude can be fitted to the available data, with the magnitudes and phase for each resonance floated freely. The choice of resonances in this model, and the fitted values of amplitude and phase, must both be optimised.

In this analysis, datasets are fitted with and generated from models using the MINT software package, which has previously been used by both the CLEO and LHCb collaborations~\cite{MattMiranda,Jonasfeasibility}. The software constructs a negative log-likelihood function from the sum of amplitudes, which is then minimised using the MINUIT~\cite{MINUIT} package.

Once a model has been fitted, it is important to quantify the quality of the fit. This is achieved by computing a $\chi^{2}$ per degree of freedom ($\chi^2_{d.o.f.}$) between binned data events and those generated with the optimised model. The binning is constructed from the signal region data in 5-dimensional phase space, requiring a minimum of 10 events per bin. Cycling through each dimension, all bins with 20 or more events are split into two approximately equal bins, until no bin contains more than 19 events. The $\chi^{2}$ per degree of freedom is given by:

\begin{equation}
\chi^{2}_{d.o.f.} = \frac{{\displaystyle\sum_{p=1}^{n} (N_{p} - N^{exp}_{p})^{2}/{N^{exp}_{p}}}}{N_{bins} - N_{fit~parameters}},
\end{equation} where $N_{p}$ is the number of observed and $N_{p}^{exp}$ is the expected number of events for bin $p$, while $N_{bin}$ is the total number of bins and $N_{fit~parameters}$ are the number of freely varied parameters in the fitted model.

\section{Model Building and Fits to Background Samples}
\subsection{Algorithmic Model Searches}
To find an optimised solution efficiently, from a pool of candidate components, genetic algorithms are frequently used. This class of algorithm treats candidate solutions (or individuals) as a sum of characteristics (genes), where these individuals can be tested under some metric for success. This metric must decide if the individual will `breed' with other individuals from the population and contribute new individuals to the next generation. This breeding mechanism can be varied but implies the construction of a new individual from a mix of genes of both parents. Often mutation is introduced randomly, either by the addition of a gene from neither parent, or modification of an existing gene.

Populations are therefore constructed from the genes of successful individuals from the previous generation, and are themselves tested for fitness. The algorithm continues like this in iterations, until some termination condition is reached. For example, reaching a maximum number of generations or a lack of improvement between generations.

Here a genetic algorithm has been designed to optimise an amplitude model by repeatedly fitting candidate models to a data sample, and using the $\chi^{2}$ per degree of freedom as a measure of model `fitness'. We begin with a partly random population of 25 models and fit these to our data sample using MINT. The reported $\chi^{2}_{d.o.f.}$ for each model is then used to rank the population, where only the top 7 will continue to `breed'. The breeding partners and number of children for each pairing is based on rank and has been designed to favour higher ranking models with more children, as seen in Table~\ref{tab:BreedingPatt}.

\begin{table}[t]
\begin{center}
\begin{tabular}{l|ccccccc}  

 Model Rank &  1st & 2nd & 3rd & 4th & 5th & 6th & 7th \\ \hline
 1st & - & 6 & 4 & 3 & 1& 1& 1\\
 2nd & 6 & - & 3 & 3 & 0 & 0 & 0\\
 3rd &  4 &  3 & - & 3 & 0 & 0 & 0\\
 4th &  3 & 3 & 3 & - & 0 & 0 & 0\\
 5th & 1 & 0& 0 &  0 & - & 0 & 0\\
 6th & 1 & 0 &0 &  0 & 0 & - & 0\\
 7th & 1 & 0 & 0 &  0 & 0 & 0 & - \\
 \hline
 
 \end{tabular}
\caption{The number of unique offspring allowed for each pairing, based on rank received by each model. For example, the top performing model and the second best will breed together and contribute six daughters for the next generation, whereas the pairing of the top model with the 7th only contributes one.}
\label{tab:BreedingPatt}
\end{center}
\end{table}

The breeding mechanism allows a 50\% chance that a component (gene) from either parents will appear in the daughter model. The chance of adding a random component is then determined by the current model size and is 1 for models less than 5 and $\frac{1}{3}$ for larger models. Additional components can then be added with a probability of $\frac{1}{3}$, i.e. a model of 5 components has a 1 in 9 chance of receiving 2 additional components. 

This is a relatively high rate of mutation, and would quickly lead to very large models, which are not considered to be beneficial. Therefore we introduce the random removal of components from models above a particular size. If the model contains between 5 and 9 components there is a 1 in 6 chance of removal, whereas models greater than 9 have a 1 in 3 chance of removal. As before, once a removal is decided (the component is chosen at random) there is an additional 1 in 3 chance of a second removal. Therefore, for models with 9 components or more, the rate of removal equals the rate of addition.

The probabilities outlined above can be varied but the reported values have been optimised from tests with simulated data.

\subsection{Physical Considerations}
This iterative approach requires an initial population of models from which to `evolve'. Two properties are desired from this base generation. First, the population must be genetically diverse such that future generations have a sizeable pool of components to choose from, and second, the original models should reflect pre-existing knowledge of likely model components.

To achieve these goals, all possible components were grouped into four basic categories, dependent on the resonances they describe. Any component with an intermediate $a_{1}(1260)$ resonance is placed in the first group, any component with a $\rho(770)$ but not an  $a_{1}(1260)$ is placed in the second. The third group contains only one component, direct non-resonant phase-space, while the fourth contains all of the remaining possibilities. The initial models are then generated by selecting a component from each group at random. The models are then compared to one another to ensure they are all unique. This grouping structure therefore ensures each model is genetically unique and contain resonances expected in the dataset, while spreading similar components throughout the population.

\subsection{Performance on Simulated Data}
\begin{figure}[htb]
\centering
\includegraphics[trim = 0mm 0mm 0mm 8mm, clip, scale = 0.7]{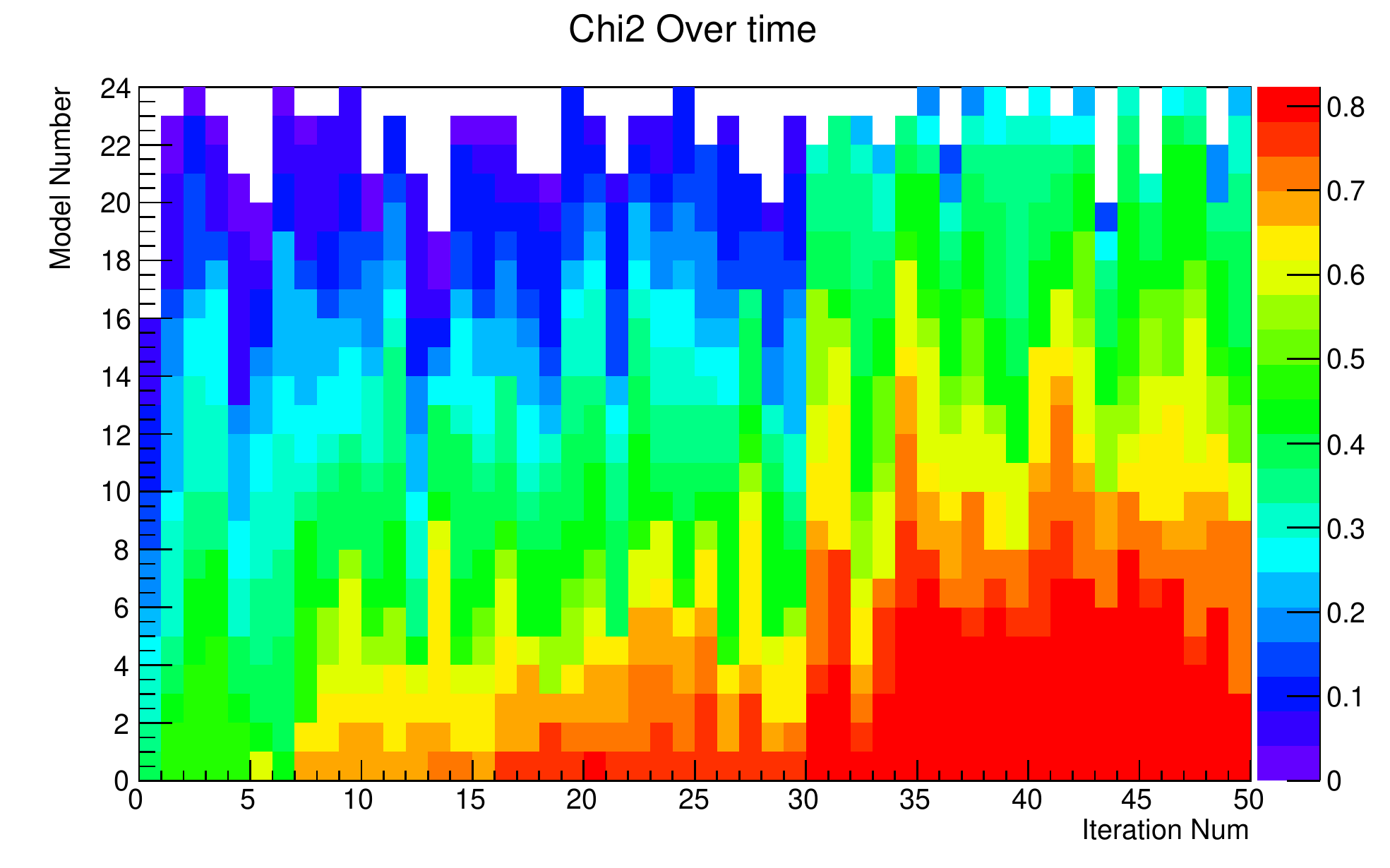}
\caption{The inverse of the $\chi^{2}$ per degree of freedom for each model, showing gradual improvement with successive iterations. Blank spaces indicate failed models, often those that exceeded their CPU time limit to fit the data.
}
\label{fig:ChiBlock}
\end{figure}

\begin{figure}[htb]
\centering
\includegraphics[trim = 20mm 22mm 0mm 20mm, clip, scale = 0.6]{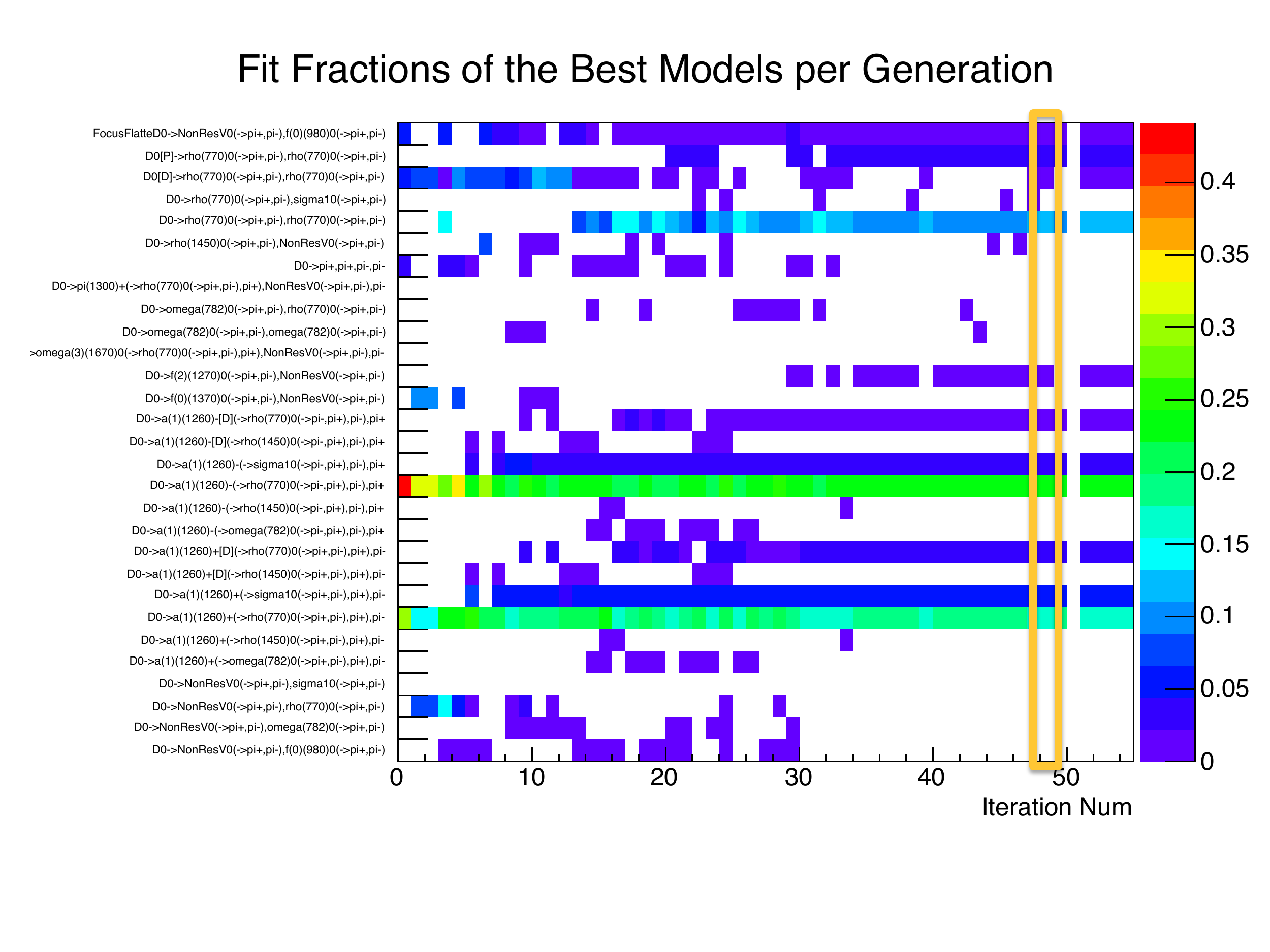}
\caption{Fit fractions for the components in the highest ranking model of each generation. Blank fields indicate the component was not present in the model. The block of fractions on the right shows the original MC model, and the best model of iteration 49, highlighted in orange) shows the same set of components and fractions.
}
\label{fig:ModelComp}
\end{figure}

To assess the performance of the algorithm, 10 000 events (comparable to the available CLEO-c dataset) were generated using MINT and an approximation of the FOCUS model. These were then processed by the algorithm. The results of this study can be seen in Figures~\ref{fig:ChiBlock} and~\ref{fig:ModelComp}. Figure~\ref{fig:ModelComp} reports the components used and their respective fit fractions for highest ranked model in each generation. The fit fraction here is defined as the amplitude and line shape integrated over all of the phase space, divided by the integrated sum of all amplitudes and line shapes. The algorithm successfully identifies the original model on the 49th iteration.

\subsection{Results from Side Band Data}
The signal region, as defined in Section~\ref{subsec:SignalandBackgroundRegions}, is expected to contain some fraction of combinatoric background. It is therefore important to describe the background distribution with an amplitude model. Then, a background P.D.F. can be added to candidate signal P.D.F.s in the expected ratio. For the purpose of fitting a background model, the two side band region datasets were merged together, forming a single background sample of approximately 4000 events. 

\begin{figure}[htb]
\centering
\includegraphics[trim = 0mm 0mm 0mm 15mm, clip, scale = 0.5]{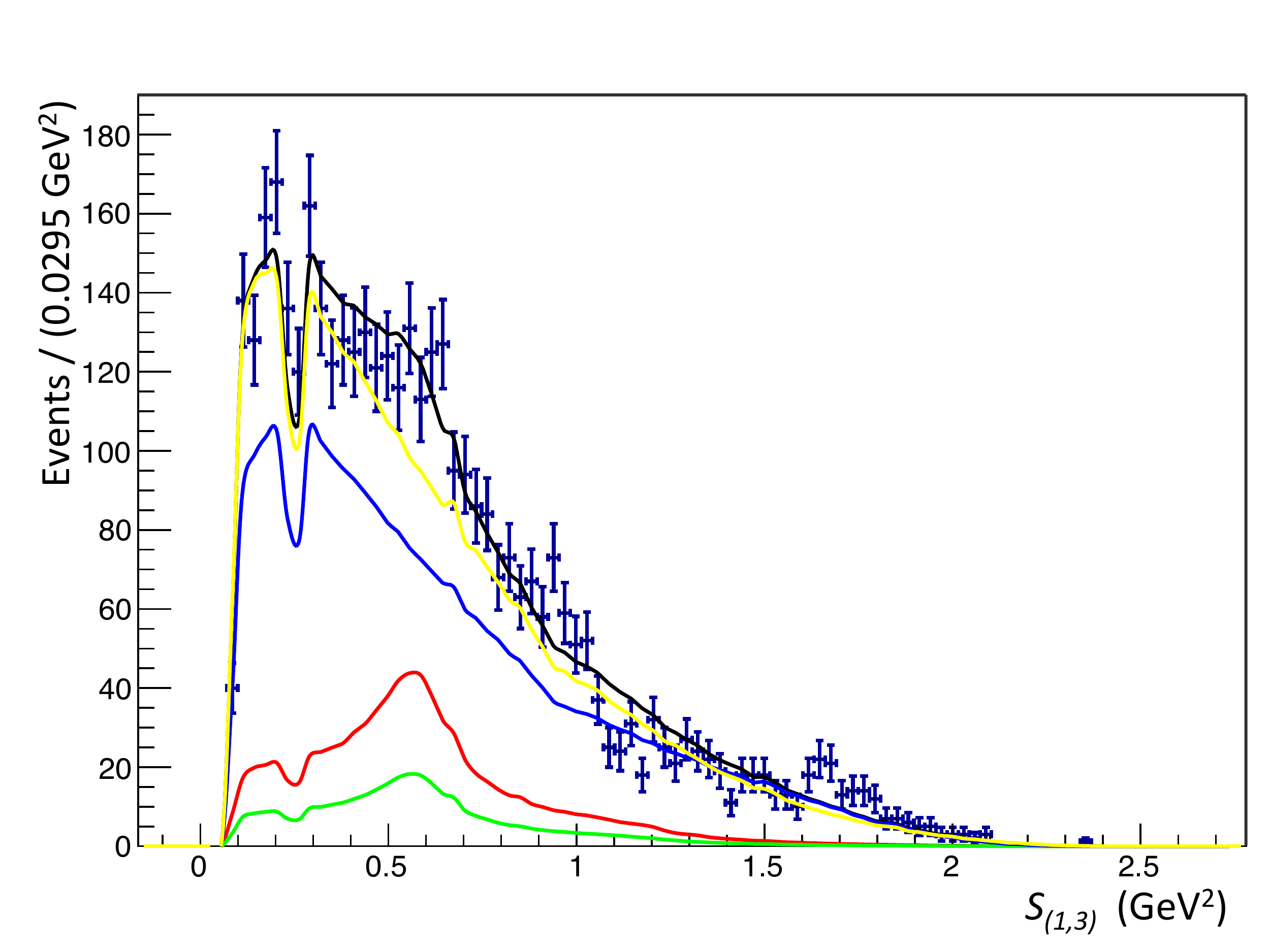}
\caption{The fit to background region data in the invariant mass of $\pi^{-}\pi^{+}$ ($s_{1,3})$, phase space projection. The total fit (black) is made from four components, as shown in Table~\ref{tab:BGModel}. Following the order of the table, the components are coloured red, green, blue and yellow respectively. The area of distributions indicate the relative fit fraction of each component in the model.}
\label{fig:BGfit}
\end{figure}

It is not expected for any resonant shapes in a background model to interfere with one another coherently. Therefore the background model can be built from amplitude components as previously described, but without any interference terms. The background sample is then fitted using the algorithm, finding the model in Table~\ref{tab:BGModel} to have the lowest $\chi^{2}_{d.o.f.}$ after 100 iterations. 
Figure~\ref{fig:BGfit} shows the contributions of the individual background model components to the final fit, in one of the 10 possible phase-space projections.

\begin{table}[t]
\begin{center}
\begin{tabular}{l|c}  

 Decay Component &  Fit Fraction \\ \hline
 $\pi^{-}, a_{1}(1260)^{+}  \rightarrow (\pi^{+}, \rho(770)^{0} \rightarrow{(\pi^{+}\pi^{-})})$  &  0.11 $\pm$ 0.12    \\
 $\pi^{+},a_{1}(1260)^{-} \rightarrow (\pi^{-}, \rho(770)^{0} \rightarrow{(\pi^{-}\pi^{+})})$ &  0.04 $\pm$ 0.11 \\ 
 $\pi^{+},\pi^{-},f_{0}(1370)^{0} \rightarrow (\pi^{+}\pi^{-})$ & 0.37 $\pm$ 0.07 \\
 $\pi^{+},\pi^{-},\pi^{+},\pi^{-}$ (Direct, non-resonant phase space) & 0.48 $\pm$ 0.09 \\

\end{tabular}
\caption{The components and relative fit fractions of the final selected background model.}
\label{tab:BGModel}
\end{center}
\end{table}

\section{Discussion}
The performance of the algorithm, at both identifying the model used to generate a simulated data sample and discovering a simple and accurate background model is encouraging. Work is now underway to prepare a final run of the algorithm to the CLEO-c signal region data, with simultaneous fits of $D^{0}~\rightarrow~\pi^{-}\pi^{+}\pi^{+}\pi^{-}$, $\overline{D^{0}}~\rightarrow~\pi^{-}\pi^{+}\pi^{+}\pi^{-}$, while accounting for the combinatoric background and the 4.5\% of mis-tagged events.

\clearpage

\Acknowledgements
We would like thank the CESR staff, CLEO collaboration and Cornell University for the provision of data, resources and support,
as well as the numerous communities behind the multiple open source software packages that we depend on. This work was supported by the U.K. Science and Technology Facilities Council.

\end{document}

%% file: charmmacros.tex
\newcommand\pubnumber{\pbnr}
\newcommand\pubdate{\today}
\def\Title#1{\begin{center} {\Large #1 } \end{center}}
\def\Author#1{\begin{center}{ \sc #1} \end{center}}

\newcommand{\OnBehalf}[1]{\sbox0{#1}\ifdim\wd0=0pt
        {}
	\else
	{\\on behalf of #1}
	\fi}
\newcommand{\SupportedBy}[1]{\sbox0{#1}\ifdim\wd0=0pt
        {}
	\else
	{\footnote{#1}}
	\fi}
\def\Address#1{\begin{center}{ \it #1} \end{center}}

\newcommand\pubblock{\includegraphics[width=5cm]{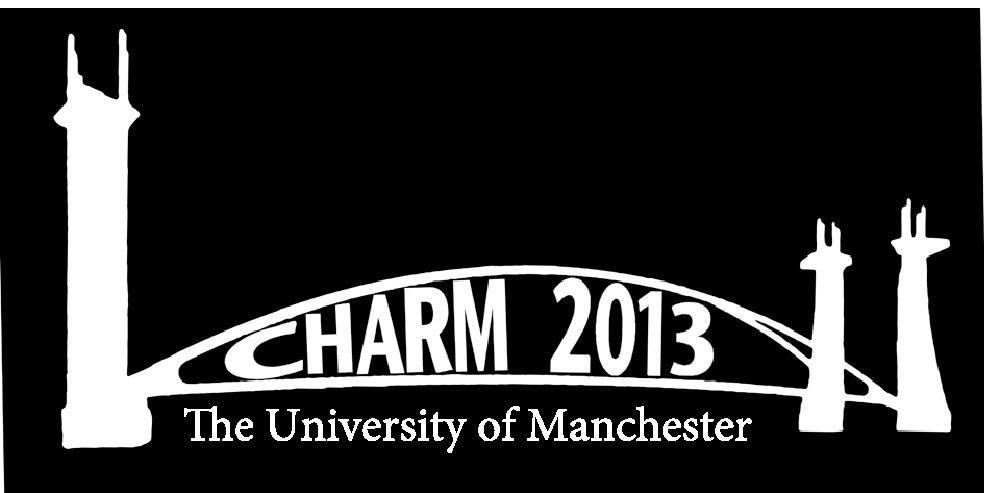}\hfill{\begin{tabular}{l} \pubnumber\\
         \pubdate  \end{tabular}}}
\newenvironment{Abstract}{\begin{quotation}  }{\end{quotation}}
\newenvironment{Presented}{\begin{quotation} \begin{center} 
             PRESENTED AT\end{center}\bigskip 
      \begin{center}\begin{large}}{\end{large}\end{center} \end{quotation}}
\def\Acknowledgements{\bigskip  \bigskip \begin{center} \begin{large}
             \bf ACKNOWLEDGEMENTS \end{large}\end{center}}
\def\venue{The 6$^{th}$ International Workshop on Charm Physics\\
(CHARM 2013)\\
Manchester, UK,  31 August -- 4 September, 2013}

\def\beq{\begin{equation}}
\def\eeq#1{\label{#1}\end{equation}}
\def\eeqn{\end{equation}}

\def\beqa{\begin{eqnarray}}
\def\eeqa#1{\label{#1}\end{eqnarray}}
\def\eeqan{\end{eqnarray}}

\let\bar=\overbar

\def\Dslash{\not{\hbox{\kern-4pt $D$}}}
\def\dslash{\not{\hbox{\kern-2pt $\del$}}}

\def\msb{{\bar{\ssstyle M \kern -1pt S}}}

